\documentclass{jfm}
\usepackage[utf8]{inputenc}

\renewcommand{\pagelimitfooter}{}
\renewcommand{\absfooterflag}{}

\newcounter{ncorresp}
\let\Oldcorresp\corresp
\renewcommand{\corresp}[1]{{\Oldcorresp{#1}}\stepcounter{ncorresp}}

\usepackage{graphicx}
\usepackage{newtxtext}
\usepackage{newtxmath}
\usepackage{natbib}
\usepackage{hyperref}
\hypersetup{
    colorlinks = true,
    urlcolor   = blue,
    citecolor  = black,
}

\newcommand{\RomanNumeralCaps}[1]
\linenumbers

\usepackage{aas_macros} 
\usepackage{mathtools} 
\usepackage{siunitx}
\usepackage{bm} 
\usepackage{booktabs} 
\usepackage{subcaption} 
\usepackage{multirow} 

\usepackage[loadonly]{enumitem}
\newlist{enuminline}{enumerate*}{1} 
\setlist[enuminline]{label=(\roman*)}

\newcommand{\parencite}{\citep}
\newcommand{\textcite}{\citet}
\newcommand{\Textcite}{\Citet}

\newcommand{\tablelinespace}{\addlinespace[5pt]} 
\newlength{\tablecolextraspace} 
\newcommand{\brokenmidrule}[2]{\cmidrule(r{\tablecolextraspace}){#1} \cmidrule{#2}}

\renewcommand{\dh}{\partial}
\renewcommand{\d}{\mathrm{d}}
\newcommand{\D}{\mathrm{D}} 
\newcommand{\mean}[1]{\mkern 1.5mu\overline{\mkern -1.5mu#1\mkern -1.5mu}\mkern 1.5mu}
\newcommand{\meanBr}[1]{\left<#1\right>}

\newcommand{\dive}{\nabla\!\cdot}
\newcommand{\curl}{\nabla\!\times\!}
\newcommand{\grad}{\nabla}
\newcommand{\Lap}{\nabla^2}

\newcommand{\defn}{\equiv}
\newcommand{\abs}[1]{\left|#1\right|}

\renewcommand{\vec}[1]{\boldsymbol{#1}}
\newcommand{\cross}{\times}
\newcommand{\Dirac}{\operatorname{\delta}}
\newcommand{\uvec}[1]{\boldsymbol{\hat{#1}}} 

\newcommand{\dyad}[1]{\overset{\text{\tiny$\bm\leftrightarrow$}}{#1}}
\newcommand{\pencil}{\textsc{Pencil}} 

\newcommand{\R}{\ensuremath{\Rey}{}} 
\newcommand{\PrM}{\ensuremath{\mbox{\textit{Pr}$_m$}}{}} 
\newcommand{\Ma}{\ensuremath{\mbox{\textit{Ma}}}} 
\newcommand{\Co}{\ensuremath{\mbox{\textit{Co}}}} 
\newcommand{\ReyTay}{\ensuremath{\mbox{\textit{Re}$_\tau$}}{}} 

\newcommand{\vecF}{\vec{F}} 
\newcommand{\vecFvisc}{\vec{F}^\text{(visc)}} 
\newcommand{\vecFlor}{\vec{F}^\text{(Lor)}} 
\newcommand{\kZE}{k_\mathrm{ZE}} 
\newcommand{\nuSmag}{\nu_\text{Smag}}
\newcommand{\CSmag}{C_\text{Smag}}
\newcommand{\dxmax}{\Delta} 
\newcommand{\tauto}{\ensuremath{\tau_\text{to}}} 
\newcommand{\LTay}{\ensuremath{L_\text{Tay}}} 
\newcommand{\Emagbykin}{\ensuremath{\widetilde{E}}} 
\newcommand{\LKol}{\ensuremath{L_\text{Kol}}} 

\newcommand{\ssdLowEtaSimsLabel}{A} 

\title{Rotational effects on the small-scale dynamo}
\author{
G. Kishore
\corresp{\email{kishoreg@iucaa.in}}
\and
Nishant K. Singh
\corresp{\email{nishant@iucaa.in}}
}
\affiliation{Inter-University Centre for Astronomy \& Astrophysics, Post Bag 4, Ganeshkhind, Pune 411 007, India}

\begin{document}

\maketitle
\addtocounter{footnote}{\value{ncorresp}}

\begin{abstract}
	Using direct numerical simulations of forced rotating turbulence, we study the effect of rotation on the growth rate and the saturation level of the small-scale dynamo.
	For slow rotation rates, increasing the rotation rate reduces both the growth rate and the saturation level.
	Once the rotation rate crosses a threshold, large-scale vortices are formed which enhance the growth rate and the saturation level.
	Below this threshold, the suppression of the small-scale dynamo with increasing rotation is explained by the fact that at scales close to, but smaller than, the forcing scale, rotating turbulence is one-dimensionalized, with the velocity component along the rotation axis being larger than the other two components.
	This is due to the rotational destabilization of vortices produced by the forcing function.
	While the rotational effect on the growth rate becomes small at high $\Rey$, the ratio of the steady-state magnetic to kinetic energies remains suppressed by up to 35\% as compared to the non-rotating case.
\end{abstract}

\section{Introduction}

Magnetic fields are ubiquitous in astrophysics, being found in stars, planets, galaxies, and even galaxy clusters \citep[for a review, see][section 2]{KanduPhysicsReports2005}.
The predominant explanation for the generation and sustenance of such fields is the turbulent dynamo, wherein a pre-existing seed magnetic field is amplified by the turbulent motion of a conducting fluid \citep{MoffattMagFieldGenBook, KrauseRadler80, KanduPhysicsReports2005, ShukurovKanduBook}.

Theoretically, dynamo mechanisms are classified into two types based on the correlation length of the generated magnetic field.
In large-scale dynamos, the correlation length of the magnetic field is much larger than that of the turbulent velocity field.
Such dynamos have been extensively studied, both analytically and numerically \citep[for a review, see][sections 6--10]{KanduPhysicsReports2005}.
In small-scale dynamos (SSD), which are the focus of this work, the correlation length of the magnetic field is of the order of or smaller than that of the turbulent velocity field.
Note that these two kinds of mechanisms can coexist in the same physical system.

The earliest analytical studies of the SSD \parencite{kazantsev1968enhancement, kulsrud92} considered a homogeneous, isotropic, incompressible, nonhelical, Gaussian, and Markovian velocity field.
Later analytical and numerical studies attempted to relax these assumptions.
For example, \textcite{vainshtein86} and \textcite{MalBol07, MalBol10} have studied the effect of helicity;
\textcite{FedSchBov14} the effect of compressibility;
\textcite{KopKisIly22} the effect of non-Gaussianity;
\textcite{bhat2014fluctuation}, and \textcite{GopSin24} the effect of a nonzero correlation time;
\textcite{SkoSquBha21} the effect of stratification;
and \textcite{singh2017enhancement} the effect of shear.

Since many systems of astrophysical interest rotate, it is natural to ask how rotation affects the SSD.
\Textcite{FavBus12} have studied the effect of rotation on SSD action in compressible (i.e.\@ non-Boussinesq) convection.
Comparing a rotating simulation with a non-rotating one, they claim to find ``weak evidence'' that the rotating cases saturate at a higher level.
Their statement rests on a comparison of time series short enough that fluctuations in the magnetic and kinetic energies are expected to affect their conclusion.
We are not aware of any other analytical or numerical studies of the effect of rotation on the SSD.

Rotation is expected to have the following effects:
\begin{enuminline}
	\item the turbulent velocity field may become anisotropic at scales larger than the Zeman scale \parencite{Zem94,BaqDav15};
	and
	\item rotation itself may generate large-scale flows (\citealp{GueHugJon17}; \citealp{BusKapMas18}; \citealp[fig.~3]{kapyla2019Lambda}).
\end{enuminline}
The effect of anisotropy on the SSD can be inferred from the work of \textcite{SkoSquBha21}, who have found that stratification suppresses the growth rate of the SSD by making the velocity field two-dimensional at large scales.
Naively, one expects that if rotation were to make the velocity field anisotropic, it would also have a similar effect.
On the other hand, as an example of the effect of large-scale flows on the SSD, one can consider the simulations by \textcite{singh2017enhancement}, who found that the growth rate of the SSD is enhanced when large-scale shear is imposed on the system.
Given the opposite directions of these two effects, it is thus not clear, a priori, if rotation enhances or suppresses the SSD.
Further, neither of these studies looked at the saturation level
(i.e.\@ the ratio of the magnetic energy to the kinetic energy in the steady state)
of the SSD.
Rather, they restricted themselves to the kinematic phase, where the magnetic field is so weak that it does not affect the velocity field.
The nonlinear phase of the SSD is of more astrophysical interest than the kinematic phase; the SSD grows on a timescale related to the eddy turnover time, much smaller than the typical ages of astrophysical objects.

In this work, we use direct numerical simulations to study the effect of rotation on the growth rate and the saturation level of the SSD.
Section \ref{section: numerical setup} describes our numerical setup.
In section \ref{section: ssd-rot}, we discuss simulations of the SSD in forced rotating turbulence.
To understand our findings, we then analyse a set of purely hydrodynamic simulations in section \ref{lsv: section: small-scale anisotropy}.
Finally, section \ref{section: conclusions} summarizes our work and discusses future directions.

\section{Numerical methods}
\label{section: numerical setup}

\subsection{Equations, domain, and boundary conditions}

We consider isothermal forced turbulence in a cubical periodic box of side $L$.
Rotation is treated in the $f$-plane approximation: spatial variation of the Coriolis force is ignored, and there is no centrifugal force.
We choose units in which the sound speed $c$, the initial (uniform) density $\rho_0$, and the magnetic permeability $\mu_0$ are 1; and $L = 2\pi$.
These choices mean that if we define $k_0 \defn 2\pi/L$, the unit for time is $\left( c k_0 \right)^{-1}$; length is $k_0^{-1}$; mass is $\rho_0 k_0^{-3}$; and magnetic field is $\sqrt{\mu_0 \rho_0 c^2 }$.
In the simulations described in section \ref{section: ssd-rot}, we solve the continuity, momentum, and induction equations:
\begin{align}
	\begin{split}
		\frac{\D \ln\rho}{\d t} ={}& - \dive\vec{u}
	\end{split}
	\\
	\begin{split}
		\rho \, \frac{\D \vec{u}}{\d t}
		={}&
		- c^2 \grad \rho
		- 2 \rho \, \vec{\Omega} \cross \vec{u}
		+ \rho \vecFvisc
		+ \rho \vecFlor
		+ \rho \vecF
	\end{split}
	\\
	\begin{split}
		\frac{\dh \vec{A} }{\dh t} ={}& \vec{u}\cross\vec{B} + \eta\Lap\vec{A}
	\end{split}
\end{align}
where
\begin{align}
	\begin{split}
		\vecFlor
		\defn{}&
		\frac{ \vec{J} \cross \vec{B} }{\rho}
	\end{split}
	\\
	\begin{split}
		\vec{B} \defn{}& \curl\vec{A}
	\end{split}
	\\
	\begin{split}
		\vec{J} \defn{}& \frac{ \curl\vec{B} }{\mu_0}
	\end{split}
	\\
	\begin{split}
		\vecFvisc
		\defn{}&
		\frac{1}{\rho} \dive{\!\left[ 2 \rho \nu \dyad{S} \right]}
		\label{eq: Fvisc newtonian}
	\end{split}
	\\
	\begin{split}
		S_{ij}
		\defn{}&
		\frac{1}{2} \left( \dh_iu_j + \dh_ju_i - \frac{2}{3} \delta_{ij}\dive\vec{u}\right)
	\end{split}
\end{align}
with
$\rho$ being the density;
$\vec{u}$ the velocity;
$\D/\d t \defn \dh/\dh t + \vec{u} \cdot \grad$ the convective derivative;
$c$ the speed of sound;
$\Omega$ the rotational rate;
$\vecF$ a forcing function;
$\vecFvisc$ the viscous force;
$\vecFlor$ the Lorentz force;
$\vec{A}$ the magnetic vector potential;
$\nu$ the kinematic viscosity;
$\mu_0$ the magnetic permeability; and
$\eta$ the magnetic diffusivity.
Later on, in section \ref{lsv: section: small-scale anisotropy}, we consider purely hydrodynamic simulations, where we solve just the continuity and momentum equations (as given above, but without the Lorentz force).
We choose a coordinate system such that $\vec{\Omega}$ is along the $z$ axis ($\vec{\Omega} = \Omega \uvec{z}$), and thus refer to the latter as the axial direction.

For all our simulations except those described in section \ref{section: effect forcing}, the forcing function is given by \parencite{HauBraMee04}
\begin{equation}
	\vecF = f_0 c \sqrt{\frac{k c}{\delta t}} \, \frac{\vec{k} \cross \uvec{e}}{\sqrt{ k^2 - \left( \vec{k} \cdot \uvec{e} \right)^2} } \cos{\!\left( \vec{k} \cdot \vec{x} + \phi \right)}
	\label{lsv: eq: force=helical}
\end{equation}
where $\delta t$ is the length of the timestep, and $f_0$ is a dimensionless factor that controls the amplitude of the forcing.
The quantities $\vec{k}$, $\uvec{e}$, and $\phi$ are randomly chosen at each timestep, subject to the following constraints:
$\vec{k}$ is chosen such that $2.5 < k/k_0 < 3.5$ (resulting in a mean wavenumber $k_f \approx 3.13 k_0$);
$\uvec{e}$ is a unit vector randomly chosen in the plane normal to $\vec{k}$;
and $-\pi \le \phi < \pi$.
Further, we choose $f_0 = \num{3e-2}$.
Since this forcing function is divergence-free, we refer to it as \emph{solenoidal} forcing.

The equations described above are solved using the \pencil{} code \parencite{Pencil2021}.\footnote{
\url{https://pencil-code.nordita.org}
}
Spatial derivatives are discretized using a sixth-order finite difference scheme, while the equations are evolved in time using a third-order Runge-Kutta method.
Upwinding is used for advection of the density, the velocity, and the magnetic vector potential \parencite[appendix B]{DobStiBra06}.
We use the same number of grid points in each direction.
In appendix \ref{grid-indep: section}, we verify that the number of grid points we choose is sufficient to resolve the phenomena we are interested in.

The initial velocity is set to zero, while the initial density is set to $\rho_0$.
Each component of the magnetic vector potential at each grid point is initially drawn from a Gaussian distribution with mean 0 and standard deviation $\num{4e-6} \sqrt{\mu_0 \rho_0} \, c / k_0$.
The timestep is chosen to keep the Courant numbers based on the advective and diffusive terms less than 0.6 and 0.25 respectively.

\subsection{Diagnostics}
To characterize the solutions to the simulations described above, we use the following dimensionless numbers.
The Reynolds number ($\Rey \defn u_\text{rms} L_f / \nu$, where $L_f \defn 2\pi/k_f$) characterizes how turbulent the flow is.
The Coriolis number ($\Co \defn \Omega L_f / u_\text{rms} $, where $u_\text{rms}$ is the root mean square value of the velocity over the entire box) expresses the importance of rotation relative to advection.
Note that this is inversely proportional to the more commonly used Rossby number.
Additionally, we also quote the Mach number ($\Ma \defn u_\text{rms} / c$).
The magnetic Prandtl number, $\PrM \defn \nu/\eta$, is an independent dimensionless number; throughout this study, we set $\PrM = 1$.

\subsection{Data analysis}

In the context of the SSD, we are interested in comparing quantities across the kinematic and saturated phases.
It is thus useful to have an automated way of computing averages over the kinematic and saturated phases given a particular simulation.

The first issue is that since we start from a fluid at rest, the velocity field takes some time to reach
a statistically
steady state under the influence of the forcing function.
We thus exclude the first five turnover times while computing averages.
The turnover time is estimated as $\tauto \defn 2\pi/\left( u_\text{rms,full} k_f\right)$, where $u_\text{rms,full}$ is the average over the entire time series.

To compute averages over the kinematic phase, we consider times after the first five turnover times, but before the volume-averaged magnetic energy reaches \num{e-2} of the volume-averaged kinetic energy.
This is further divided into five chunks, the average over each of which is treated as an independent realization.
The average of the quantity of interest is then estimated by averaging the chunked averages, while the associated error is estimated as $\sigma/\sqrt{n}$, where $\sigma$ is the standard deviation of the chunked averages, and $n$ is the number of chunks.
In particular, the growth rate is estimated by fitting a straight line to the time series of the logarithm of the volume averaged magnetic energy in each chunk.

The beginning of the saturated phase is estimated to be the point where the extrapolated kinematic fit to the volume-averaged magnetic energy exceeds the time average (excluding the first five turnover times) of the volume-averaged kinetic energy.
As in the kinematic phase, errors in quantities pertaining to the saturated phase are estimated by dividing it into five chunks.

\section{The effect of rotation on the small-scale dynamo}
\label{section: ssd-rot}

\begin{table}
	\centering
	\begin{tabular}{
		c
		@{\hskip\tablecolextraspace}
		S[table-format=1.3]
		S[table-format=1.1e-1, exponent-mode=scientific, table-align-exponent=false]
		S[table-format=1.1e-1, exponent-mode=scientific, table-align-exponent=false]
		r
		@{\hskip\tablecolextraspace}
		S[table-format=4.0, round-mode=places, round-precision=0]
		S[table-format=1.2, round-mode=places, round-precision=2]
		S[table-format=1.3, round-mode=places, round-precision=3]
		@{\hskip\tablecolextraspace}
		S[table-format=1.4(2), round-mode=uncertainty]
		S[table-format=1.4(2), round-mode=uncertainty]
		}
		Name
		& {$\Omega/(c k_0)$} &   {$\eta k_0/c$} &   {$f_0$} &   {$N_x$} &   {$\R$} &   {$\Co$} &   {$\Ma$} & {$\gamma\tauto$}                          & {$\widetilde{E}$}   \\
	\brokenmidrule{1-5}{6-10} \\
		\multirow{11}{*}{A}
		& 0     &          0.001   &     0.03  &       128 &  308.416 &  0        &  0.153843 & 0.17863892714208654+-0.015221090099949779 & 0.08121491435185317+-0.0032368306502754828  \\
		&                0.025 &          0.001   &     0.03  &       128 &  308.856 &  0.340185 &  0.154063 & 0.18409931497297466+-0.012253798258403321 & 0.0737955144752195+-0.0032416448820100333   \\
		&                0.05  &          0.001   &     0.03  &       128 &  309.749 &  0.675227 &  0.154508 & 0.1701280286259391+-0.010444949821012066  & 0.06993592384740525+-0.0022926733259789394  \\
		&                0.075 &          0.001   &     0.03  &       128 &  311.354 &  1.00206  &  0.155309 & 0.15704034795080482+-0.015007677752718964 & 0.056127533962237665+-0.0021714987787532277 \\
		&                0.1   &          0.001   &     0.03  &       128 &  313.856 &  1.32395  &  0.156557 & 0.15330254882373426+-0.012185732718705974 & 0.05298511872105298+-0.0013014159362479388  \\
		&                0.125 &          0.001   &     0.03  &       128 &  316.573 &  1.62599  &  0.157912 & 0.13170113440030007+-0.006294016758043014 & 0.04395964145516298+-0.00388609035148426    \\
		&                0.15  &          0.001   &     0.03  &       128 &  320.608 &  1.90915  &  0.159925 & 0.10273886429551064+-0.006866859705284846 & 0.03207652049996222+-0.0017087121672872685  \\
		&                0.175 &          0.001   &     0.03  &       128 &  326.2   &  2.19054  &  0.162714 & 0.09595443503581536+-0.008091908207807098 & 0.02826662991854704+-0.0019444747231614585  \\
		&                0.2   &          0.001   &     0.03  &       128 &  337.506 &  2.4259   &  0.168354 & 0.099683881801993+-0.006939249410830119   & 0.033700328618919154+-0.0029840706363046554 \\
		&                0.3   &          0.001   &     0.03  &       128 &  450.957 &  2.93597  &  0.224945 & 0.11885248897454027+-0.017524769589535277 & 0.045965402604426844+-0.0030243524079748402 \\
		&                0.5   &          0.001   &     0.03  &       128 &  516.33  &  4.62335  &  0.257554 & 0.17263550415992326+-0.02075701552003558  & 0.076785730654961+-0.007261253879878731     \\
	\tablelinespace
		\multirow{7}{*}{B}
		& 0     &          0.0005  &     0.026 &       256 &  585.898 &  0        &  0.146128 & 0.39936755105291183+-0.011953565135447711 & 0.15060251752534393+-0.002618864829972787   \\
		&                0.025 &          0.0005  &     0.026 &       256 &  585.869 &  0.37626  &  0.146121 & 0.40748448808134907+-0.01606829858366977  & 0.14610575014100108+-0.0038058620847319868  \\
		&                0.05  &          0.0005  &     0.026 &       256 &  588.101 &  0.742508 &  0.146677 & 0.3885285470921196+-0.01735879650097008   & 0.13122014468621046+-0.0013038424533632514  \\
		&                0.075 &          0.0005  &     0.026 &       256 &  591.923 &  1.1068   &  0.147631 & 0.3676357274440193+-0.019142184352204994  & 0.13535875295291613+-0.0034732129692484942  \\
		&                0.1   &          0.0005  &     0.026 &       256 &  598.471 &  1.44898  &  0.149264 & 0.3422301851731368+-0.00793079906322684   & 0.11820742845455572+-0.002506040315079652   \\
		&                0.125 &          0.0005  &     0.026 &       256 &  600.459 &  1.78453  &  0.14976  & 0.32158866929679447+-0.013936342091717749 & 0.1053074799776351+-0.003478983550690626    \\
		&                0.15  &          0.0005  &     0.026 &       256 &  609.496 &  2.10247  &  0.152013 & 0.3210201340524281+-0.009642806150536212  & 0.09055062749761694+-0.0013589706631596844  \\
	\tablelinespace
		\multirow{4}{*}{C}
		& 0     &          0.00025 &     0.026 &       512 & 1201.73  &  0        &  0.149861 & 0.6592858508225499+-0.02216407667262593   & 0.21616592432330095+-0.008683520281109377   \\
		&                0.05  &          0.00025 &     0.026 &       512 & 1197.57  &  0.747175 &  0.149342 & 0.6175273043440147+-0.027860730177325134  & 0.19941346348924888+-0.007473908773878046   \\
		&                0.1   &          0.00025 &     0.026 &       512 & 1218.69  &  1.45332  &  0.151976 & 0.6038014056094334+-0.023133015187393958  & 0.17186800215733727+-0.009497305559389803   \\
		&                0.15  &          0.00025 &     0.026 &       512 & 1247.47  &  2.10279  &  0.155565 & 0.5620301781039139+-0.01935748295002464   & 0.14028225340859596+-0.0036508761438087205  \\
	\end{tabular}
	\caption{
		Parameter values and diagnostics (computed by averaging over the kinematic phase) for the magnetohydrodynamic simulations.
		In all these simulations, $\nu=\eta$.
		For $\Omega \ge 0.2 \, c k_0$ (where a large-scale vortex is formed), $u_\text{rms}$ does not reach a statistically steady state within the kinematic phase; this affects the quoted values of $\Rey$, $\Co$, and $\Ma$.
		}
	\label{table: all ssd sims}
\end{table}

\begin{figure}
	\centering
	\includegraphics{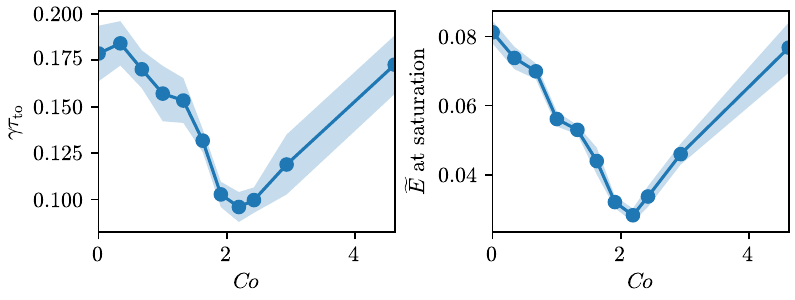}
	\caption{
		Left: growth rate (scaled by the turnover time) of the magnetic energy for different Coriolis numbers in the set of simulations with $\eta = \num{e-3} \, c/k_0$ (marked \ssdLowEtaSimsLabel{} in table \ref{table: all ssd sims}).
		The shaded region represents the estimated error in the growth rate.
		Right: the ratio of magnetic to kinetic energies in the saturated phase for different rotation rates in the same set of simulations.
		}
	\label{B3.SSDaniso.rot: fig: gr and bsat vs Omega}
\end{figure}
\begin{figure}
	\centering
	\includegraphics{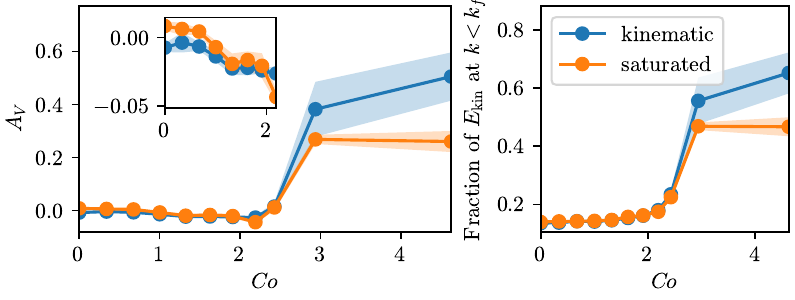}
	\caption{
		Left: level of anisotropy induced by rotation in the same simulations as in figure \ref{B3.SSDaniso.rot: fig: gr and bsat vs Omega}.
		The inset on the left shows a zoomed-in view of the low-$\Omega$ region.
		Right: at high rotation rates, a large fraction of the kinetic energy resides at spatial scales larger than the forcing scale.
		}
	\label{B3.SSDaniso.rot: fig: AV and large-scale-ekin vs Omega}
\end{figure}
\begin{figure}
	\centering
	\includegraphics{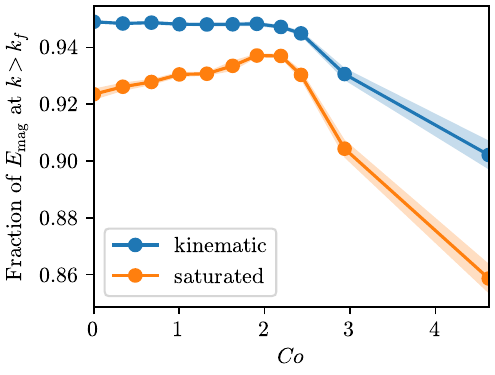}
	\caption{
		The fraction of the magnetic energy at spatial scales smaller than the forcing scale, as a function of the Coriolis number, in the same simulations as in figure \ref{B3.SSDaniso.rot: fig: gr and bsat vs Omega}.
		}
	\label{B3.SSDaniso.rot: fig: small-scale emag vs Omega}
\end{figure}
\begin{figure}
	\centering
	\includegraphics{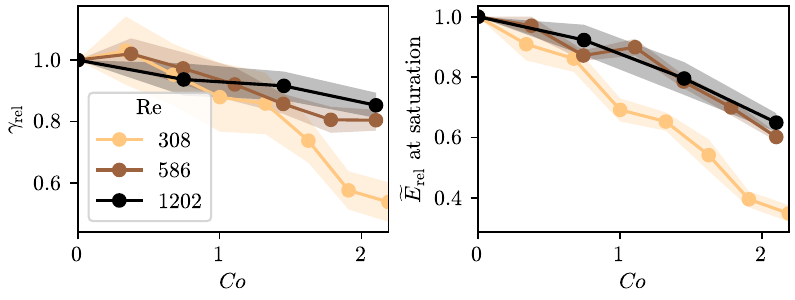}
	\caption{
		Kinematic growth rate (left) and the ratio of magnetic to kinetic energies in the saturated phase (right) as a function of the Coriolis number for different Reynolds numbers.
		Each set of simulations (table \ref{table: all ssd sims}) has been labelled by the Reynolds number of its non-rotating case.
		The subscript `rel' denotes that the quantity plotted has been normalized by the corresponding value in the absence of rotation.
		}
	\label{fig: ssd gr and sat frac vs Co different Re}
\end{figure}

First, we consider a set of simulations where the rotation rate varies, but all other parameters are kept the same (the set \ssdLowEtaSimsLabel{} in table \ref{table: all ssd sims}).
For notational convenience, we use $\Emagbykin$ to denote the ratio of the magnetic energy to the kinetic energy.
Figure \ref{B3.SSDaniso.rot: fig: gr and bsat vs Omega} shows that for
$\Co \lesssim 2$,
both the growth rate and the saturation level of the magnetic field decrease as the rotational rate increases.
However, at higher rotational rates, the growth rate and the saturation level increase with rotation.
Note from table \ref{table: all ssd sims} that $\Rey$ increases monotonically with the rotation rate; this does not explain the non-monotonic trend in the growth rate.

Since the SSD is known to be affected by anisotropization of the velocity field, let us now check if similar trends are seen in the level of anisotropy of the latter.
Following \citeauthor{Kap19} (\citeyear[eq.~17]{Kap19}; \citeyear[eq.~30]{kapyla2019Lambda}), we quantify the anisotropy of the flow by
\begin{equation}
	A_V \defn \frac{ \meanBr{u_x^2} + \meanBr{u_y^2} - 2 \meanBr{u_z^2} }{ \meanBr{u^2} }
	\label{B3.SSDaniso: eq: A_V defn}
\end{equation}
where $\meanBr{\Box}$ denotes the volume average of $\Box$.
Note that $-2 \le A_V \le 1$, and $A_V=0$ if the turbulence is isotropic.
Recall that we have chosen a coordinate system such that the rotation vector is along the $z$ axis.

The left panel of figure \ref{B3.SSDaniso.rot: fig: AV and large-scale-ekin vs Omega} shows that the flow is almost isotropic below $\Co \approx 2$.
For faster rotation rates, the flow becomes highly anisotropic, with the components of the velocity field perpendicular to the rotation axis being larger than that along the rotation axis.
The right panel of figure \ref{B3.SSDaniso.rot: fig: AV and large-scale-ekin vs Omega} shows that in this regime, the fraction of the kinetic energy\footnote{
This was actually calculated from the velocity power spectrum.
Since these simulations are at low Mach number, we do not expect our results to be affected by neglecting density fluctuations.
} above the forcing scale is much larger than in the slowly rotating regime.
The phenomenon we observe here is referred to as the \emph{large-scale vortex instability}, and is known to be capable of exciting a large-scale dynamo \citep{GueHugJon17, BusKapMas18}.
Indeed, figure \ref{B3.SSDaniso.rot: fig: small-scale emag vs Omega} shows that
when $\Co \gtrsim 2$,
the fraction of the magnetic energy at scales larger than the forcing scale increases with the rotation rate, suggesting that what we observe in this regime is better described as a large-scale dynamo coexisting with the small-scale dynamo.

Going back to the left panel of figure \ref{B3.SSDaniso.rot: fig: AV and large-scale-ekin vs Omega}, note that for
$\Co \lesssim 2$,
$A_V$ becomes more negative as the rotation rate increases (both in the kinematic and saturated phases).
This corresponds to the component of the velocity field along the rotation axis being larger than the other two components,
and may be connected to the decrease in the growth rate and the saturation level of the SSD with increasing rotation.

Figure \ref{fig: ssd gr and sat frac vs Co different Re} shows the growth rate and the saturation level of the magnetic field as a function of the Coriolis number, but this time for different values of the Reynolds number.
We restrict ourselves to rotation rates where the large-scale vortex instability is not excited.
Since both the growth rate and the saturation level of the magnetic field are known to be functions of $\Rey$ and $\PrM$ (\citealp[fig.~14]{schekochihin2004ssdSim}; \citealp{FedSchBov14}), we express these quantities in the rotating simulations as fractions of their values in otherwise identical non-rotating simulations.

At the highest $\Rey$ considered, the growth rate of the SSD does not vary much with rotation, being suppressed by only around 15\% at $\Co \approx 2$.
Further, the strength of this effect seems to decrease as $\Rey$ increases, suggesting
that rotation does not affect the growth rate of the SSD in the astrophysically relevant limit of high $\Rey$.
However, the steady-state ratio of magnetic to kinetic energies behaves somewhat differently: for a particular rotational rate, at the two highest values of $\Rey$ considered, this ratio remains the same fraction of its non-rotating value.
This raises the possibility that the rotational effect on the saturation level of the magnetic field remains significant at high $\Rey$.
Quantitatively, at the highest $\Rey$ considered, the ratio of magnetic to kinetic energies is suppressed by about 35\% at $\Co \approx 2$ as compared to its non-rotating value.

The suppression of the SSD which we observe with increasing rotation in the absence of the large-scale vortex instability disagrees with the claim made by \citet{FavBus12}.
However, we note that their suggestion that rotation enhances the saturation level of the SSD relies on a time series short enough to be significantly affected by stochastic fluctuations.

\section{Generation of anisotropy in rotating forced turbulence}
\label{lsv: section: small-scale anisotropy}

\subsection{Scale-dependence of rotationally generated anisotropy}
\label{section: lsv irrelevant}

\begin{table}
	\centering
	\begin{tabular}{
		S[table-format=1.3]
		l
		S[table-format=1e-1, exponent-mode=scientific]
		l
		S[table-format=1.1e-1, exponent-mode=scientific, table-align-exponent=false]
		r
		@{\hskip\tablecolextraspace}
		S[table-format=1.2, round-mode=places, round-precision=2]
		S[table-format=4.0, round-mode=places, round-precision=0]
		S[table-format=4.0, round-mode=places, round-precision=0]
		S[table-format=2.1, round-mode=places, round-precision=1]
		S[table-format=1.3, round-mode=places, round-precision=3]
		}
		{$\Omega/(c k_0)$} & viscosity   & {$\nu k_0/c$}   & forcing      &   {$f_0$} &   {$N_x$} &   {$\Co$} & {$\R$}             & {$\ReyTay$}        &   {$L_f/L_\mathrm{Tay}$} &   {$\Ma$} \\
	\brokenmidrule{1-6}{7-11}
		0.05 & normal      & 0.005           & solenoidal   &      0.03 &        64 &  0.89987  & 44.66196397100552  & 14.819758809998765 &                  3.01368 &  0.111391 \\
		0.15 & normal      & 0.005           & solenoidal   &      0.03 &        64 &  2.65785  & 45.36370033811715  & 15.13663306601827  &                  2.99695 &  0.113141 \\
		0.25 & normal      & 0.005           & solenoidal   &      0.03 &        64 &  4.4422   & 45.23653902224963  & 15.308596403762836 &                  2.95498 &  0.112824 \\
		0.5  & normal      & 0.005           & solenoidal   &      0.03 &        64 &  8.72992  & 46.037049108927505 & 15.656476765065852 &                  2.94045 &  0.11482  \\
	\tablelinespace
		0.05 & normal      & 0.0005          & solenoidal   &      0.03 &       256 &  0.621651 & 646.5032689182598  & 98.167656891089    &                  6.58571 &  0.161243 \\
		0.15 & normal      & 0.0005          & solenoidal   &      0.03 &       256 &  1.80993  & 666.1580314548637  & 105.84642264262493 &                  6.29363 &  0.166145 \\
		0.2  & normal      & 0.0005          & solenoidal   &      0.03 &       256 &  2.33754  & 687.7298445652266  & 112.36534478853439 &                  6.12048 &  0.171526 \\
		0.25 & normal      & 0.0005          & solenoidal   &      0.03 &       256 &  2.79894  & 717.9487078024645  & 122.60787962456772 &                  5.85565 &  0.179062 \\
	\tablelinespace
		0.25 & normal      & 0.0002          & solenoidal   &      0.03 &       256 &  2.69834  & 1861.7904634646218 & 213.0422803191013  &                  8.73907 &  0.185739 \\
		0.25 & Smagorinsky & {}              & solenoidal   &      0.03 &       256 &  2.70381  & {}                 & {}                 &                 12.5484  &  0.185363 \\
		0.25 & Smagorinsky & {}              & solenoidal   &      0.03 &       512 &  2.67863  & {}                 & {}                 &                 19.3986  &  0.187105 \\
	\tablelinespace
		0.25 & normal      & 0.0005          & irrotational &      0.85 &       256 &  2.70571  & 810.5993729669993  & 55.719544411965124 &                 14.5478  &  0.193516 \\
	\end{tabular}
	\caption{
		Parameter values and diagnostics for the hydrodynamic simulations.
		}
	\label{table: all hydro sims}
\end{table}

\begin{figure}
	\begin{subfigure}{\textwidth}
		\centering
		\caption{
			$\nu = \num{5e-4} \, c/k_0$
			}
		\label{turb.lsv.sim: fig: AV vs k for nu 5e-4}
		\includegraphics{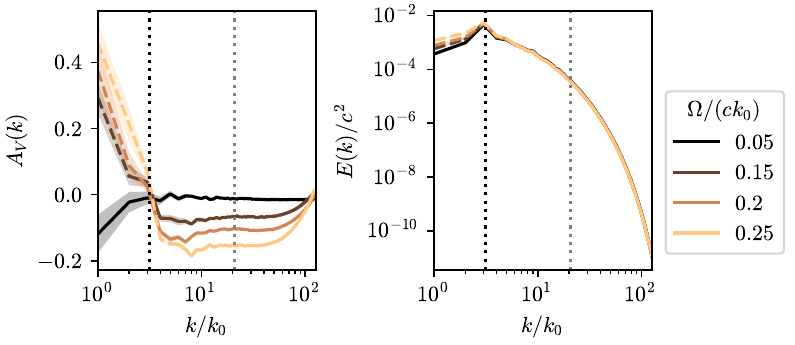}
	\end{subfigure}
	
	\begin{subfigure}{\textwidth}
		\centering
		\caption{
			$\nu = \num{5e-3} \, c/k_0$
			}
		\label{turb.lsv.sim: fig: AV vs k for nu 5e-3}
		\includegraphics{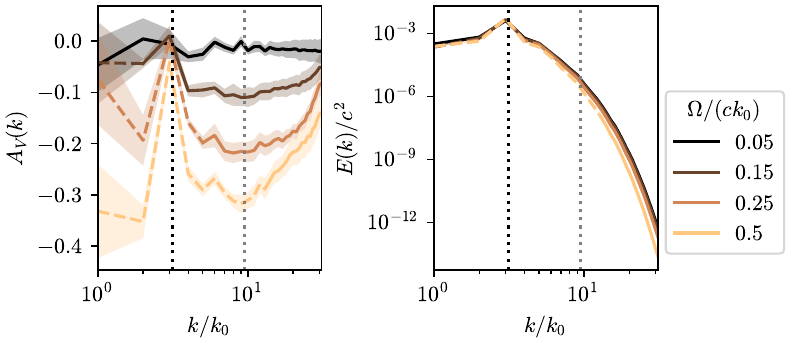}
	\end{subfigure}
	
	\caption{
		Variation of $A_V(k)$ and the specific kinetic energy spectrum with the rotation rate for two different values of $\nu$.
		The black vertical dotted line marks the forcing wavenumber, while the grey vertical dotted line marks that corresponding to the Taylor microscale (equation \ref{eq: LTay defn}) in the case with the slowest rotation.
		In each case, the portion of the spectrum at wavenumbers smaller than the Zeman wavenumber (defined in equation \ref{turb.lsv: eq: k_ZE defn}) is marked by a dashed line, while a solid line is used for larger wavenumbers.
		}
	\label{turb.lsv.sim: fig: AV vs k}
\end{figure}

To understand why rotation affects the small-scale dynamo, we now study the anisotropy of the velocity field at different spatial scales in a set of purely hydrodynamic simulations (relevant parameters and diagnostics are given in table \ref{table: all hydro sims}).
Following \textcite[eq.~27]{Kap19}, we quantify scale-dependent anisotropy by the spectral analogue of $A_V$:
\begin{equation}
	A_V(k) \defn \frac{ E_x(k) + E_y(k) - 2 E_z(k) }{ E(k) }
	\label{T.aniso: eq: A_V(k) defn}
\end{equation}
where, for example, $E_x(k)$ is the scalar 3D power spectrum of $u_x$.
Appendix \ref{section: anisotropy length scales vs components} discusses why we do not use the more conventional characterization based on parallel and perpendicular length scales.

Figure \ref{turb.lsv.sim: fig: AV vs k for nu 5e-4} shows $A_V(k)$ for different values of $\Omega$ in simulations with the same value of $\nu$.
Consistent with what we saw earlier, we find that at wavenumbers smaller than the forcing wavenumber, $A_V(k)$ attains large, positive values as $\Omega$ increases.
However, we also find that at wavenumbers larger than the forcing wavenumber, $A_V(k)$ attains \emph{negative} values.
Figure \ref{turb.lsv.sim: fig: AV vs k for nu 5e-3} shows a series of simulations with a larger value of $\nu$, where no large-scale vortex is formed.
We still find that $A_V$ is negative at wavenumbers larger than the forcing wavenumber, suggesting that the effect at play is independent of the large-scale vortex instability.

In simulations of the decay of initially isotropic, rotating, incompressible turbulence, \textcite[fig.~8b]{BarMetLes94} observed a similar strengthening of the axial component of the velocity field at small scales.
They explained this by the fact that vortices with vorticities (in a rotating frame) antiparallel to the rotational axis become unstable, according to the Rayleigh criterion \citep[section 66]{chandrasekhar1961hydromagnetic}, when the rotational rate is of the same order as their vorticity.
The resulting instability leads to three-dimensionalization of such flow structures.
Since this mechanism tends to drain energy from flows perpendicular to the rotation axis, the net result is that the axial component of the velocity field becomes larger than the other two components.
Relative strengthening of the axial component of the velocity field is also predicted by the EDQNM calculations of \Textcite[fig.~7]{CamJac89}.\footnote{
\Textcite[p.~14]{BarMetLes94} mention that \textcite{TeiDan87} observed this effect in direct numerical simulations; lacking access to the study by \citeauthor{TeiDan87}, we have not been able to verify this.
}

In rotating turbulence, the Zeman wavenumber \parencite[eq.~6]{Zem94},
\begin{equation}
	\kZE \defn \frac{ \Omega^{3/2} }{ \epsilon^{1/2} }
	\label{turb.lsv: eq: k_ZE defn}
\end{equation}
is usually thought to separate rotationally influenced small wavenumbers from isotropic large wavenumbers \parencite[e.g.][pp.~2,14]{BaqDav15}; the idea is that when $k > \kZE$, the turnover time of the velocity field is smaller than the rotational period, leading to rotational effects being unimportant.
In figures \ref{turb.lsv.sim: fig: AV vs k for nu 5e-4} and \ref{turb.lsv.sim: fig: AV vs k for nu 5e-3}, for each value of $\Omega$, the Zeman wavenumber has been indicated by using a dashed line for $k < \kZE$, and a solid line for $k > \kZE$.
In figure \ref{turb.lsv.sim: fig: AV vs k for nu 5e-3}, we clearly see that the position of the Zeman wavenumber does not affect the formation of the negative-$A_V$ subrange at $k > k_f$.

\subsection{Return to isotropy at small scales for large \texorpdfstring{$\Rey$}{Re}}
\label{section: large Re}

\begin{figure}
	\centering
	\includegraphics{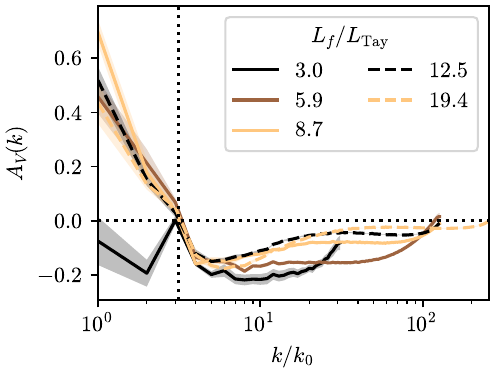}
	\caption{
		Effect of the position of the dissipative scale on $A_V(k)$ in a set of simulations with the same rotation rate ($\Omega = 0.25 \, c k_0$).
		Solid lines denote simulations with normal viscosity (constant $\nu$), while dashed lines denote simulations with Smagorinsky viscosity.
		The black vertical dotted line marks the forcing wavenumber.
		}
	\label{fig: AV vs k effect Re}
\end{figure}

As long as $\R$ is large enough, one expects turbulence to become isotropic at wavenumbers much larger than the Zeman and integral wavenumbers.
However, simulations that resolve the large range of wavenumbers required to see such an effect are computationally expensive.
To get a hint of what happens at higher $\R$ than those discussed above, we additionally consider large-eddy simulations in which the viscous force is still given by equation \ref{eq: Fvisc newtonian}, but the kinematic viscosity, $\nu$, is replaced by the Smagorinsky viscosity \citep[section 12.2.8]{lesieur2008turbulence}:
\begin{equation}
	\nuSmag
	\defn
	\left(\CSmag \dxmax\right)^2 \sqrt{2 S_{ij}S_{ij}}
\end{equation}
where $\dxmax$ is the grid spacing, and $\CSmag = 0.2$ (the recommended value of $\CSmag$ is between 0.1 and 0.2).
The parameters for these simulations are given in table \ref{table: all hydro sims} along with some diagnostics.

We define the Taylor microscale, which describes the typical length scale associated with velocity gradients in the flow (\citealp{Tay35}; \citealp[eq.~6.64]{lesieur2008turbulence}), as
\begin{equation}
	\LTay
	\defn
	\sqrt{\frac{\meanBr{u^2}}{\meanBr{u_{x,x}^2} + \meanBr{u_{y,y}^2} + \meanBr{u_{z,z}^2}}}
	\label{eq: LTay defn}
\end{equation}
where, e.g., $u_{x,x} \defn \dh u_x/\dh x$.
In homogeneous and isotropic turbulence, it is thought of as the length scale below which viscous effects start becoming important.
Figure \ref{fig: AV vs k effect Re} shows that as the Taylor microscale gets further and further from the integral scale, the flow indeed becomes more isotropic at large wavenumbers.
The subrange of constant $A_V(k)$ in figure \ref{turb.lsv.sim: fig: AV vs k} is then explained by the inertial range being too small to allow a complete return to isotropy.

\subsection{The effect of the forcing function}
\label{section: effect forcing}

\begin{figure}
	\centering
	\includegraphics{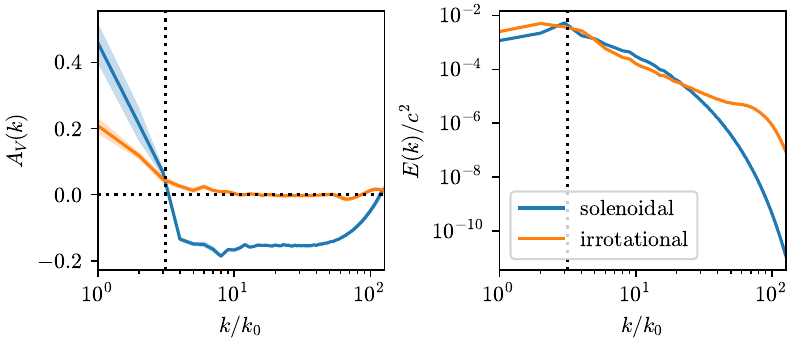}
	\caption{
		Comparison of $A_V(k)$ and $E(k)$ in a pair of simulations which differ only in the type of forcing function used.
		The black vertical dotted line marks $k_f$.
		}
	\label{fig: AV vs k effect forcing type}
\end{figure}

At this stage, we hypothesize that $A_V(k)$ becoming negative at scales just below the forcing scale is associated with the decay of vortices injected by the forcing function.
To verify this hypothesis, we consider a simulation with an irrotational forcing function.
Instead of equation \ref{lsv: eq: force=helical}, we use \parencite{MeeBra06}
\begin{equation}
	\vecF
	=
	\nabla \phi
	\,,\quad
	\phi \defn f_0 c \sqrt{\frac{c R}{\delta t}} \, \exp{\!\left( - \frac{ \left( \vec{x} - \vec{a} \right)^2 }{R^2} \right)}
	\label{lsv: eq: force=gaussianpot}
\end{equation}
where $\vec{a}$ is randomly chosen at each timestep, and $f_0$ is dimensionless.
\Citet[eq.~8]{MeeBra06} have shown that the resulting velocity spectrum peaks near the wavenumber $k_f \defn 2/R$; for easy comparison with the solenoidally forced simulations (for which $k_f \approx 3.13$), we choose $R=2/3$.
The last row of table \ref{table: all hydro sims} gives the parameters for this simulation along with some diagnostics.

Figure \ref{fig: AV vs k effect forcing type} shows that with the irrotational forcing function described above, $A_V(k)$ is close to zero at wavenumbers larger than that of the forcing scale.
Recalling our earlier discussion of the work of \textcite{BarMetLes94}, we conclude that the one-dimensionalization of turbulence below the forcing scale is due to rotational effects on vortices injected by the forcing function.
Note that large-scale vortices are formed at low wavenumbers even in the irrotationally forced case; this is likely due to vorticity generated in the dissipative range undergoing an inverse cascade.

\section{Discussion and conclusions}
\label{section: conclusions}

Rotational effects on vortices in a solenoidally forced turbulent flow can cause one-dimensionalization of the velocity field at scales smaller than the integral scale, with the component along the rotational axis becoming significantly larger than the other two components.
We have found that this reduces both the growth rate and the saturation level of the SSD in solenoidally forced rotating turbulence.
While the rotational effect on the growth rate becomes small at the highest $\Rey$ considered, the saturation level is less sensitive to $\Rey$ (still being suppressed by up to 35\%).

A possible explanation for why the growth rate and the saturation level behave differently is the fact that the correlation length of the magnetic field generated by the SSD is larger in the saturated phase than in the kinematic phase \citep[e.g.][table 1]{SetBusShu20}.
This would mean that at high $\Rey$, SSD action takes place closer to the rotationally affected scales in the nonlinear phase than in the kinematic phase.
Alternatively, vorticity generated by the Lorentz force \citep{BraSca25} may play a role.

Since the effect we describe is absent in irrotationally forced simulations, one would naively argue that our findings are not relevant to, e.g., turbulence forced by supernovae in the interstellar medium.
However, recall that the isothermal equation of state we have used precludes the existence of baroclinic torques.
In supersonic turbulence with a more realistic equation of state, vorticity is indeed generated by baroclinic torques \citep{DelBra11}.
Whether rotation still affects the small-scale dynamo in such systems depends on the scales at which baroclinic torques generate vortices.
We are not aware of any studies that answer this question, which should be dealt with in future work.

More generally, our finding --- that when the anisotropy of the velocity field is scale-dependent, the kinematic and nonlinear phases of the SSD are affected differently --- suggests that the nonlinear phase of the SSD in stratified or shear-driven turbulence should also be studied and compared with the known behaviour in the respective kinematic regimes.

\backsection[Acknowledgements]{
We acknowledge use of the Pegasus computing facility at IUCAA.
We thank Alexandra Elbakyan for facilitating access to scientific literature.
}

\backsection[Funding]{
This research received no specific grant from any funding agency, commercial or not-for-profit sectors.
}

\backsection[Data availability]{
The input files and output (volume averages and power spectra) for all the simulations used in this work, along with the scripts used to generate the figures, are available on Zenodo \citep{zenodo_ssd_rot_lsv}.
}

\backsection[Software]{
Pencil \citep{Pencil2021};
SciPy \citep{scipy2020};
NumPy \citep{numpy2020};
and Matplotlib \citep{matplotlib2007}.
}

\backsection[Declaration of interests]{
The authors report no conflict of interest.
}

\backsection[Author ORCID]{
GK, \url{https://orcid.org/0000-0003-2620-790X}; 
NS, \url{https://orcid.org/0000-0001-6097-688X}
}

\backsection[Author contributions]{
GK conceptualized the research and performed the simulations.
GK and NS interpreted the results and wrote the paper.
}

\bibliographystyle{jfm}
\bibliography{refs.bib}

\appendix

\section{Grid-dependence tests}
\label{grid-indep: section}

\begin{figure}
	\centering
	\includegraphics{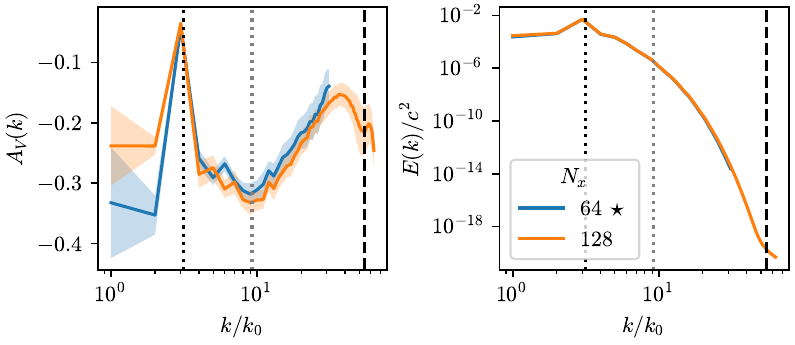}
	\caption{
		Resolution-dependence of $A_V(k)$ and $E(k)$ in the case with $\nu = \num{5e-3} \, c/k_0$ and $\Omega = 0.5 \, c k_0$.
		The star in the legend marks the number of grid points (along a single direction) that we have used in our study.
		The vertical lines mark the forcing wavenumber (black dotted), the Taylor-scale wavenumber (grey dotted), and the Kolmogorov wavenumber (black dashed).
		}
	\label{grid-indep-hydro: fig: nu 5e-3 om 0.5}
\end{figure}
\begin{figure}
	\centering
	\includegraphics{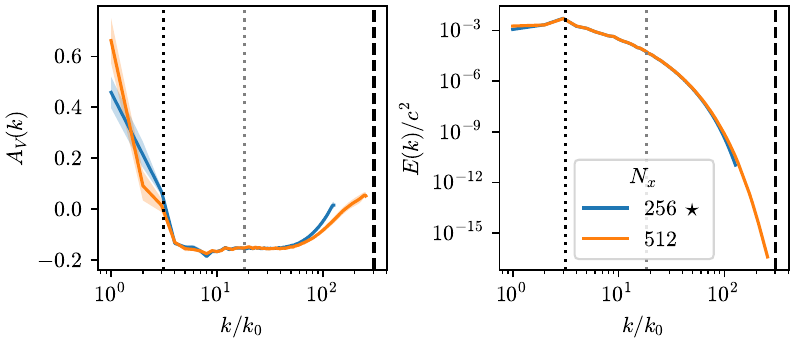}
	\caption{
		Similar to figure \ref{grid-indep-hydro: fig: nu 5e-3 om 0.5}, but for $\nu = \num{5e-4} \, c/k_0$ and $\Omega = 0.25 \, c k_0$.
		}
	\label{grid-indep-hydro: fig: nu 5e-4 om 0.25}
\end{figure}
\begin{figure}
	\centering
	\includegraphics{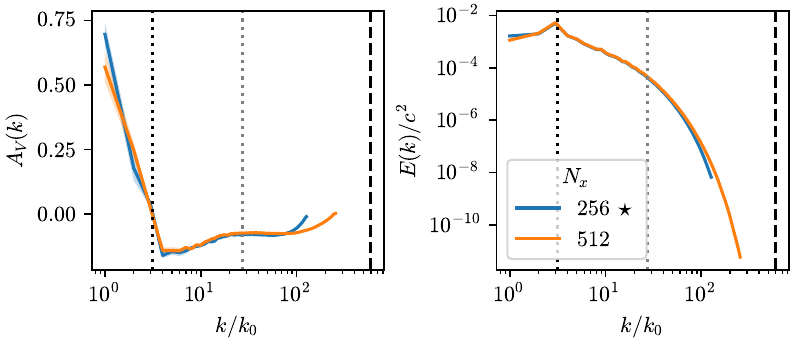}
	\caption{
		Similar to figure \ref{grid-indep-hydro: fig: nu 5e-3 om 0.5}, but for $\nu = \num{2e-4} \, c/k_0$ and $\Omega = 0.25 \, c k_0$.
		}
	\label{grid-indep-hydro: fig: nu 2e-4 om 0.25}
\end{figure}

Figures \ref{grid-indep-hydro: fig: nu 5e-3 om 0.5}, \ref{grid-indep-hydro: fig: nu 5e-4 om 0.25}, and \ref{grid-indep-hydro: fig: nu 2e-4 om 0.25} show the resolution-dependence of $A_V(k)$ and $E(k)$ for the three different values of $\nu$ used in our solenoidally forced hydrodynamic simulations (with the rotational rate in each case being the most extreme value considered).
While $A_V(k)$ and $E(k)$ at the lowest wavenumbers still show some dependence on the grid resolution, they are both reliable at $k>k_f$.
For the sake of completeness, we have also indicated the wavenumber corresponding to the Kolmogorov length scale, the latter being defined as $\LKol \defn \left( \mean{\rho} \nu^3 /\epsilon \right)^{1/4}$ with $\epsilon \defn \meanBr{2 \nu \rho S_{ij}S_{ij}}$ \parencite[e.g.][eq.~1.3]{davidson2004turbulence}.

\section{On alternative methods of quantifying anisotropy}
\label{section: anisotropy length scales vs components}

The conventional method of characterizing the scale-dependence of the anisotropy of turbulence is to compute the velocity power spectrum by restricting the wavevectors which are summed over.
Assuming the direction of anisotropy is $\uvec{z}$ and denoting
$\Phi_{ii}(\vec{k}) \defn \meanBr{u_i(\vec{k}) \, u_i^*(\vec{k})}$,
one defines \parencite[see][eqs.~1,2]{BaqDav15}
\begin{align}
	\begin{split}
		E(k)
		\defn{}&
		\frac{1}{2} \int\d\vec{k} \Dirac(\abs{\vec{k}} - k) \, \Phi_{ii}(\vec{k})
		\label{T.aniso: eq: E(k) defn}
	\end{split}
	\\
	\begin{split}
		E_\parallel(k)
		\defn{}&
		\frac{1}{2} \int\d\vec{k} \Dirac(\abs{\vec{k}\cdot\uvec{z}} - k) \, \Phi_{ii}(\vec{k})
		\label{T.aniso: eq: E_par defn}
	\end{split}
	\\
	\begin{split}
		E_\perp(k)
		\defn{}&
		\frac{1}{2} \int\d\vec{k} \Dirac(\abs{\vec{k}\cross\uvec{z}} - k) \, \Phi_{ii}(\vec{k})
		\label{T.aniso: eq: E_perp defn}
	\end{split}
\end{align}
Note that these are normalized such that \parencite[eq.~3]{BaqDav15}
\begin{equation}
	\int_0^\infty \d k \, E(k)
	=
	\int_0^\infty \d k \, E_\parallel(k)
	=
	\int_0^\infty \d k \, E_\perp(k)
	=
	\frac{1}{2} \meanBr{u^2}
\end{equation}
The quantities $E_\perp(k)$ and $E_\parallel(k)$\footnote{
These are conventionally denoted as $E(k_\perp)$ and $E(k_\parallel)$ respectively.
} seem to be preferred in theoretical work \parencite[e.g.][]{GolSri95,NazSch11}.
This appendix explains why, in this study, we quantify anisotropy using $A_V(k)$ (defined in equation \ref{T.aniso: eq: A_V(k) defn}) instead.

If the turbulence is homogeneous, isotropic, and nonhelical, we can write (see \citealp[eq.~5.84]{lesieur2008turbulence})
\begin{equation}
	\Phi_{ii}(\vec{k}) = \frac{E(\abs{\vec{k}})}{2\pi \abs{\vec{k}}^2}
\end{equation}
Plugging this into equation \ref{T.aniso: eq: E_par defn}, we find
\begin{align}
	\begin{split}
		E_\parallel(a)
		={}&
		\frac{1}{4\pi} \int\d\vec{k} \Dirac(\abs{\vec{k}\cdot\uvec{z}} - a) \, \frac{E(\abs{\vec{k}})}{\abs{\vec{k}}^2}
	\end{split}
	\\
	\begin{split}
		={}&
		\frac{1}{2\pi} \int\d\vec{k}_\perp \, \frac{ E{\!\left(\sqrt{a^2 + \abs{\vec{k}_\perp}^2} \right)} }{ a^2 + \abs{\vec{k}_\perp}^2 }
	\end{split}
	\\
	\begin{split}
		={}&
		\int_0^\infty k_\perp \d k_\perp \, \frac{ E{\!\left(\sqrt{a^2 + k_\perp^2} \right)} }{ a^2 + k_\perp^2 }
	\end{split}
	\\
	\begin{split}
		={}&
		\int_0^\infty \d k_\perp \, \frac{ k_\perp }{ a^2 + k_\perp^2 } \, E{\!\left(\sqrt{a^2 + k_\perp^2} \right)}
	\end{split}
	\\
	\begin{split}
		={}&
		\int_{a}^\infty \d k \, \frac{ E(k) }{ k }
		\label{T.aniso: eq: E_par E relation homo iso}
	\end{split}
\end{align}
which agrees with the relation derived by \textcite[eq.~8.37c]{davidson2004turbulence}.
Similarly, from equation \ref{T.aniso: eq: E_perp defn}, we find
\begin{align}
	\begin{split}
		E_\perp(a)
		={}&
		\frac{1}{4\pi} \int\d\vec{k} \Dirac(\abs{\vec{k}\cross\uvec{z}} - a) \, \frac{E(\abs{\vec{k}})}{ \abs{\vec{k}}^2}
	\end{split}
	\\
	\begin{split}
		={}&
		\frac{1}{4\pi} \int_{-\infty}^\infty \d k_\parallel \int \d\vec{k}_\perp \Dirac(\abs{\vec{k}_\perp} - a) \, \frac{ E{\!\left(\sqrt{k_\parallel^2 + \abs{\vec{k}_\perp}^2} \right)} }{ k_\parallel^2 + \abs{\vec{k}_\perp}^2 }
	\end{split}
	\\
	\begin{split}
		={}&
		\int_0^\infty \d k_\parallel \int_0^\infty \d k_\perp \, k_\perp \Dirac(k_\perp  - a) \, \frac{ E{\!\left(\sqrt{k_\parallel^2 + k_\perp^2} \right)} }{ k_\parallel^2 + k_\perp^2 }
	\end{split}
	\\
	\begin{split}
		={}&
		\int_0^\infty \d k_\parallel \, \frac{a}{ k_\parallel^2 + a^2 } \, E{\!\left(\sqrt{k_\parallel^2 + a^2} \right)}
	\end{split}
	\\
	\begin{split}
		={}&
		\int_a^\infty \d k \, \frac{a\,  E(k) }{k \sqrt{k^2 - a^2}}
		\label{T.aniso: eq: E_perp E relation homo iso}
	\end{split}
\end{align}
We see that even when the turbulence is homogeneous, isotropic, and nonhelical, one cannot expect $E_\parallel(k) = E_\perp(k)$; it is thus difficult to use these to quantify departures from isotropy.

Consistent with what we saw above, \textcite[sections 8.1.5,8.1.9]{davidson2004turbulence} notes that in general, 1D (e.g. $E_\parallel(k)$) or 2D (e.g. $E_\perp(k)$) spectra cannot be interpreted in the same way as 3D spectra (e.g. $E(k)$).
The point is that, for example, fixing the value of $k_\parallel$ (which is just the projection of $\vec{k}$ onto $\uvec{z}$) does not fix the value of $\abs{\vec{k}}$, and thus $E_\parallel(k_\parallel)$ depends on all modes for which $\abs{\vec{k}} \ge k_\parallel$.
\Citeauthor{davidson2004turbulence} refers to this as \emph{aliasing}.

\end{document}